\documentclass[amstex]{aa}
\usepackage{graphics}
\begin{document}

\thesaurus{12         
              (02.01.2;  
               02.18.8;  
               02.18.5;  
	       02.18.6)} 
  \title{Comptonization of photons in advection dominated accretion flows}
  \subtitle{Monte Carlo approach}

   \author{ A. Kurpiewski 
           \and M. Jaroszy\'nski }

\institute{Warsaw University Observatory, Al. Ujazdowskie 4,
 00--478 Warsaw, Poland
}

\date{Received   /Accepted }

\maketitle

\begin{abstract}

We study the formation of the spectra in advection dominated accretion
flows (ADAFs) around Kerr black holes. We use a Monte Carlo approach and
fully general relativistic treatment to follow the paths of individual
photons and model their scattering with mildly relativistic, thermal
electrons of the two temperature plasma present in the flow. We are
mostly interested in the dependence of the spectra on the black hole
angular momentum, and we find that the influence of the black hole
rotation rate on the flow structure has an impact on the resulting
spectra. The flows around the fast rotating holes produce
 relatively harder spectra. This property of the models should be taken
into account when modeling the individual sources and the population of
inefficiently accreting black holes in the Universe.

\end{abstract}

\section{Introduction}

The Advection Dominated Accretion Flows (ADAFs) have been reviewed
by a number of authors, most recently by Narayan, Mahadevan, \& Quataert 
(\cite{NMQ}), Kato, Fukue, \& Mineshige (\cite{kato}), and Lasota
(\cite{jpl98}). The ADAFs represent a class of optically thin solutions of
the accretion flow, which radiate so inefficiently, that almost all the
heat dissipated inside the fluid is subsequently advected toward the black
hole horizon. Since the cooling of matter is negligible, the equations of
fluid dynamics are independent of the equations describing the emission,
absorption and scattering of radiation. This allows one to address these two
topics separately. In this paper we are concerned mostly with the radiation
processes inside the flow.

The fully relativistic dynamics of ADAFs has been described and some
solutions have been obtained in Lasota (\cite{jpl94}), 
Abramowicz et al. (\cite{ACGL}, hereafter ACGL), 
Abramowicz, Lanza \& Percival (\cite{ALP}), 
Peitz \& Appl (\cite{peitz}), 
Jaroszy\'{n}ski \& Kurpiewski (\cite{paperI}, hereafter Paper I), 
Gammie \& Popham (\cite{gammie98}), 
and Popham \& Gammie (\cite{popham98}). 
The most extensive survey of the parameter
space is probably presented by Popham \& Gammie (\cite{popham98}), where
the dependence of the flow on the black hole spin $a$, the viscosity
parameter $\alpha$, the gas adiabatic index and on the advection parameter
$f$ (where $f=1$ means fully advective flow without any cooling) is
investigated. This work shows rather strong dependence of the flow
characteristics on parameters mentioned, especially on the black hole spin.
The dependence on advection parameter is also substantial, but for a narrow
range of $f$, which can represent flows with negligible cooling 
($0.9 \le f \le 1$) one can use the $f=1$ solutions. Similarly,
the solutions depend strongly on the viscosity parameter $\alpha$, but for
the limited range of this parameter ($\alpha \ge 0.01$), which is more
physically relevant (Balbus, Gammie \& Hawley \cite{balbus95}), the
differences between the solutions are not dramatic, in particular the
topology of the isobars is the same. 

In our calculations we use the solutions of the equations describing the
flow dynamics presented in Paper I. 
All our models use $f=1$ and $\alpha=0.1$. Both values are
representative of the physically relevant ADAFs. The black hole spin, we
consider, is limited to the three values ($a=0$, $0.5$, and $0.9$).  

The main purpose of this paper is a self-consistent
treatment of photon Comp\-to\-ni\-za\-tion in a two temperature plasma of
an ADAF solution. To do so we need a 3D distribution of matter density,
velocity and temperature, while the standard solutions give only the
vertically averaged quantities as measured at the equator. At this point it
is important to choose the relevant method of vertical averaging and we
follow Abramowicz et al. (\cite{ALP}), Quataert \& Narayan (\cite{QN}) and 
Paper I in choosing averaging on spheres (and not on cylinders). In this
approach the matter distribution in space is limited by the centrifugal
forces barrier and does not resemble the infinite isothermal atmosphere,
obtained in the alternative approach.

We calculate the spectrum of photons leaving the flow. The photons
originate in brems\-strahlung and thermal synchrotron processes, which can be
described locally. The scattering (if any) can take place at any point of
the photon trajectory inside matter distribution and is nonlocal. We
simulate this process using Monte Carlo approach. Our simulation of photon
Comp\-to\-ni\-za\-tion is quite standard, but the flow, where the processes
take place is rather complicated. Since the optical depth for scattering is
low, a photon can travel to a distant part of the fluid before undergoing
any interaction. The relative motion of the fluid elements, where
consecutive interactions of photons with matter take place, can be
substantial. Also, the photons traveling in the densest parts of the flow,
near the horizon, can be deflected by the gravitational field. This and
other relativistic effects in photon motion are included in our study.

The main aim of our investigation is the self-consistent treatment of
Comp\-to\-ni\-za\-tion. Most non-analytical thermal Comp\-to\-ni\-za\-tion
models use iterative methods of solving the kinetic equations
(e.g. Poutanen \& Svensson \cite{poutanen}). This method has been adopted
by Narayan, Barret \& McClintock (\cite{NBC}) to calculate the spectrum of
soft X-ray transient source V404 Cyg in which, they believe, ADAF exists.
In our calculations we use Monte Carlo method which has been described
in most detail by Pozdnyakov, Sobol \& Sunyaev (\cite{PSS}) 
and G\'orecki \& Wilczewski (\cite{gorecki}). This method allows us to
follow only one photon at a time and we are not able to take into account
the creation of $e^+e^-$ pairs. However as Bj{\"o}rnsson et
al. (\cite{bjorn96}) and Kusunose \& Mineshige (\cite{kusunose96}) showed,
the role of $e^+e^-$ pairs in ADAFs is not significant. 

In the next Section we briefly characterize the model of ADAF. The
method of Monte Carlo Comp\-to\-ni\-za\-tion is presented in Sec.~3. In
Sec.~4 we present the results of calculations, showing the ADAFs spectra.
The discussion and conclusions follow in the last Section.

\section{The 3D description of the fluid}

\subsection{The accretion flow}

We investigate the stationary flow of matter and the propagation of light
in the gravitational field of a rotating Kerr black hole 
using the Boyer-Lindquist coordinates $t$, $\phi$, $r$, $\theta$ and the 
metric components $g_{ab}(r,\theta)$ as given by Bardeen (\cite{Bard}).
We follow the $(-,+,+,+)$ signature convention. 
We use geometrical units, so the speed of light
$c \equiv 1$ and the mass of the hole $M \equiv 1$. 
The Kerr parameter $a$ ($0 \le a <1$) gives
the black hole angular momentum in geometrical units. We use the
Einstein summation convention, where needed, and a semicolon for the
covariant derivative. The normalization of four velocity with the chosen
metric signature reads $u^a u_a=-1$.

The system of equations we use follows in general ACGL. Since we 
are neglecting the cooling processes at this stage, and treat the 
accretion flow in the disk approximation, two components of velocity
and the speed of sound $c_{\mathrm{S}}$ given as functions of radius, fully
describe the dynamics. We use the angular velocity $\Omega$ and the 
physical radial velocity $V$ as measured by locally nonrotating observers
as main kinematic variables (compare ACGL).
After the {\it vertical averaging} the velocity perpendicular to the 
equatorial plane is neglected ($u^\theta \equiv 0$)   and other
components of the four velocity are given as:
\begin{equation}
u^t={1 \over \sqrt{1-V^2} 
\sqrt{-g_{tt}-2 \Omega g_{t\phi}-\Omega^2 g_{\phi\phi}}}
\end{equation}
\begin{equation}
u^\phi=\Omega u^t
\end{equation}
\begin{equation}
u^r={1 \over \sqrt{g_{rr}}}{V \over \sqrt{1-V^2}}~~.
\end{equation}

The above velocity components have to be known only at the equatorial plane
and its close vicinity if one needs to obtain a system of equations
describing an ADAF in the slim disk approximation (Abramowicz et
al. \cite{maa88}). To describe the matter distribution in space (also far
from the equatorial plane) one has to introduce further specifications of
the velocity and angular momentum distribution, since only the averaged
values enter the equations. We assume that the poloidal velocity component
($u^\theta$) is absent and that the radial velocity component $V$ depends
on the radius only:  
\begin{equation}
V=V(r) ~~~~~~u^\theta \equiv 0
\end{equation}
(The BL velocity component $u^r$ depends on $\theta$ through $g_{rr}$ -
compare Eq.3)

The choice of angular momentum distribution is less obvious. Close to the
horizon, where the velocities are highest and possible kinematic effects
most important, the specific angular momentum $\ell \equiv -u_\phi/u_t$ is
approximately constant (Paper I, ACGL, Peitz \& Appl, \cite{peitz}). 
We assume $\ell$ to be exactly constant on spheres:
\begin{equation}
\ell = \ell(r)
\end{equation}
Since the metric components do depend on $\theta$ the angular velocity is
not constant on spheres:
\begin{equation}
\Omega
={g^{\phi t}u_t+g^{\phi\phi}u_\phi \over g^{tt}u_t+g^{t\phi}u_\phi}
={g^{\phi t}-\ell g^{\phi\phi} \over g^{tt}-\ell g^{t\phi}}
\end{equation}
The other kinematic quantities are given as:
\begin{equation}
u_t={1 \over \sqrt{1-V^2} 
\sqrt{-g^{tt}+2 \ell g^{t\phi}-\ell^2 g^{\phi\phi}}}
\end{equation}
\begin{equation}
u_\phi=-\ell u_t
\end{equation}

We assume that the accreting plasma contains small scale,
isotropically tangled magnetic field resulting from magnetohydrodynamical
instability (Balbus et al. \cite{balbus95}). Hence we write the total
pressure as 
\begin{equation}
p=p_\mathrm{g}+p_\mathrm{m}
~~~~p_\mathrm{g}= \beta p
~~~~p_\mathrm{m} \equiv {B^2 \over 8\pi} = (1-\beta) p
\end{equation}
where $p_\mathrm{g}$ is the gas pressure, $p_\mathrm{m}$ - the 
magnetic pressure, $B$ - the magnetic field,  and $\beta$ is a constant
parameter. The pressure, the rest mass density $\rho_0$ and the sound
velocity $c_\mathrm{S}$ are related:
\begin{equation}
p=\rho_0 c_\mathrm{S}^2
\end{equation}
In our calculations we use a two temperature plasma with a small amount
of magnetic field to represent the matter properties. Thus the ion
pressure dominates, the gas is non-relativistic and the specific enthalpy
$\mu$ is given
as: 
\begin{equation}
\mu \equiv {\epsilon+p \over \rho_0}=1+{5 \over 2}c_\mathrm{S}^2
\end{equation}
where $\epsilon$ is the total (rest mass plus thermal) energy density.

In the ADAF set of equations only the vertically averaged  sound speed
is used. It is in spirit of our approximations to postulate:
\begin{equation}
c_\mathrm{S}=c_\mathrm{S}(r)
\end{equation}
which means that the gas is isothermal on spheres. 
The electron temperature $T_\mathrm{e}$ 
is much lower than the ion temperature, so
its influence on the equation of state and the structure of the flow can be
neglected.

Our detailed (but approximate) description of the fluid kinematics makes it
possible to find the density dependence on the angular coordinate $\theta$.
The set of equations
describing ADAF contains viscosity terms, which are included in the energy
equation, but neglected in the mechanical equilibrium equations (ACGL, 
Peitz \& Appl \cite{peitz}, Paper I). Thus it is sufficient to use the
ideal fluid energy momentum tensor:
\begin{equation}
T^a_b=(\epsilon + p)u^a u_b + p \delta^a_b 
\end{equation}
The $\theta$ conservation equation, $T^a_{\theta;a}=0$ reads:
\begin{equation}
u^a u_{\theta;a} = {-p_{,\theta} \over \epsilon + p}
\end{equation}
After some algebra we obtain:
$$
{1 \over 1-V^2}(\ln u_t)_{,\theta}
-{1 \over 2}{V^2 \over 1-V^2}(ln(r^2+a^2\cos^2 \theta))_{,\theta}~~~~~~
$$
\begin{equation}
~~~~~~~~~~~~~~~~~~~~~~~~~~~~~~~~~~~~~
= -{c_\mathrm{S}^2 \over \mu}(\ln \rho_0)_{,\theta}
\end{equation}
Since $V$ does not depend on $\theta$, the LHS of the equation is a full
gradient of a quantity which can be called a potential $\psi$. This implies
the solution for the density:
\begin{equation}
\rho_0(\theta)=\rho_{0,\mathrm{eq}}\exp\left(-{\mu(\psi(\theta)-\psi_\mathrm{eq})
\over c_\mathrm{S}^2}\right)
\end{equation}
where the $\rho_{0,\mathrm{eq}}$, $\psi_\mathrm{eq}$ denote the values measured
at the equatorial plane. 

The constant sound speed on the spheres may suggest that the exponential
atmosphere never ends. For the rotating configuration there is, however, an
infinite potential barrier close to the rotation axis, where $u_t 
\rightarrow \infty$ and $\rho_0 \rightarrow 0$. Thus the vicinity of the
rotation axis is empty and the density falls steeply down near this
region. In this respect the ADAF solutions are similar to the so called 
thick accretion disks (Abramowicz, Jaroszy{\'n}ski, \& Sikora \cite{AJS78}).
Both have empty funnels around their rotation axes.

The mass flow through the $r=const$ surface can be calculated as:
\begin{equation}
{\dot M}=-\int_0^\pi \mathrm{d}\theta \mathrm{d}\phi 
\sqrt{-g} \rho_0(\theta) u^r
\end{equation}
\begin{equation}
\equiv 2\pi \sqrt{V^2 \over 1-V^2} \int_0^\pi \mathrm{d}\theta 
\sqrt{(g_{t\phi}^2-g_{tt}g_{\phi\phi})g_{\theta\theta}}\rho_0(\theta) 
\end{equation}
Combining the last equation and the formula for the $\theta$ dependence
of the density we obtain the equatorial value of the density.

\subsection{Two temperature plasma}

While modeling the dynamics of ADAF we neglect the heat transfer and all
the radiation processes, assuming that only a small part of the total energy
generated by viscous processes can be affected by them. Now we are going to
model the radiation processes. 

We assume that the whole energy dissipated is transferred to
the ions. The ions heat the electrons via Coulomb collisions and the
electrons lose their energy by the synchrotron, bremsstrahlung and
inverse Compton cooling processes. In the ADAFs thermalization time-scale
greatly exceeds the dynamical time-scale and the plasma remains two temperature.
We find the electron and ion temperatures, $T_\mathrm{e}(r,\theta)$
and $T_\mathrm{i}(r,\theta)$, self-consistently using the equation of state 
\begin{equation}
p_\mathrm{g}=\beta p=\frac{\rho
kT_\mathrm{i}}{\mu_\mathrm{i}m_\mathrm{H}}+
\frac{\rho kT_\mathrm{e}}{\mu_\mathrm{e}m_\mathrm{H}}
\label{state}
\end{equation}
where $\mu_\mathrm{i}=1.29$ and $\mu_\mathrm{e}=1.18$ are effective molecular
weights of the ions and electrons, and the condition of thermal equilibrium
applied locally
\begin{equation}
q^\mathrm{+}=q^\mathrm{-}_\mathrm{br}+q^\mathrm{-}_\mathrm{br,C}
+q^\mathrm{-}_\mathrm{S}+q^\mathrm{-}_\mathrm{S,C}
\label{equilib}
\end{equation}
where $q^\mathrm{+}$ is the rate of Coulomb heating of electrons by
ions (e.g. Mahadevan \cite{mahadevan}), $q^\mathrm{-}_\mathrm{S}$ and
$q^\mathrm{-}_\mathrm{br}$ are the synchrotron and brems\-strahlung cooling
rates, and $q^\mathrm{-}_\mathrm{S,C}$ and $q^\mathrm{-}_\mathrm{br,C}$ 
are the Compton cooling rate of synchrotron and brems\-strah\-lung photons,
respectively.

The calculation of the synchrotron cooling rate is somewhat complicated 
because the optical depth to absorption (Self Synchrotron Absorption) for
majority of the synchrotron photons is high.
In our approach we assume that a photon can either escape
from the medium carrying out its energy and taking part in cooling
process, or be absorbed very close to the point of its original
emission and not contributing to the cooling. In reality all photons
carry energy and some are absorbed far from the emission point
transferring energy to distant portions of the fluid. Our assumption
neglects the heat transport between different volume elements, but makes
calculations doable.

To find the probability of a photon escape from given location we follow
$N \sim 10^2$ rays with directions randomly chosen at the frame
comoving with the fluid. The optical depth along a ray measured at the
frequency $\nu$ is 
\begin{equation}
\tau_\nu=\int_0^{l_\infty}
\frac{\epsilon_\mathrm{S}(\nu)}{4\pi B_{T_\mathrm{e}}(\nu)}\mathrm{d}l
\end{equation}
where $\epsilon_\mathrm{S}$ is the synchrotron emissivity, 
$B_{T_\mathrm{e}}$ is the
Planck function corresponding to the electron temperature
$T_\mathrm{e}$, and the integration is over the proper distance,
$l_\infty$ corresponding to the point on the fluid boundary.
Since the integrand in the above formula is the
function of the electron temperature we must assume the approximate 
distribution of the electron temperature in the ADAF. As a first
approximation we use our results from Paper I based on the approach of
Narayan \& Yi \cite{NYi}.
Averaging one gets the probability of escape:
\begin{equation}
\mathrm{e}^{-\tau_{\nu}} =
{1 \over N}~\sum_{i=1}^N~~\mathrm{e}^{-\tau_\nu^{(i)}}
\label{tau}
\end{equation}
Since we assume that cooling is provided only by the escaping photons,
we have:
\begin{equation}
q^\mathrm{-}_\mathrm{S}=\int_0^\infty~~\epsilon_\mathrm{S}(\nu)~
\mathrm{e}^{-\tau_\nu}~\mathrm{d}\nu
\end{equation}
for the synchrotron cooling. 
We adopt the expressions for synchrotron emissivity
from Pacholczyk (\cite{pacholczyk}) and Mahadevan, Narayan \& Yi 
(\cite{MNY}).

For the brems\-strahlung cooling the solution is straightforward.
The absorption of low frequency brems\-strahlung photons has no
practical
meaning, so we adopt their frequency integrated emission as cooling rate
\begin{equation}
q^\mathrm{-}_\mathrm{br}=\int_0^\infty~\epsilon_\mathrm{br}(\nu)~\mathrm{d}\nu
\end{equation}
We take the expression for brems\-strahlung cooling rate from Stepney
\& Guilbert (\cite{stepney}).

Finally we take the cooling by Comptonization of both synchrotron and
brems\-strahlung photons using the formulae of Esin et al. (\cite{esin}).
The mean optical depth to Compton scattering is calculated in the same 
manner as above.
Solving the equations \ref{state} and \ref{equilib} we get the electron
temperature and the ion temperature.

\subsection{Spectra neglecting Comptonization}

As a by-product of the calculations of the previous subsections one can
obtain the spectrum of the model, which would be valid if the 
Comp\-to\-ni\-za\-tion were unimportant. For sufficiently low frequencies
a photon is rather absorbed than scattered, so both approaches, neglecting
and including Comp\-to\-ni\-za\-tion,  should give similar results in this
regime. This gives a chance of a self check of the simulations.

Calculation of many rays sent from a given fluid element can also be used to
find the contribution of this element to the total luminosity of the
configuration as seen by a distant observer. If the frequency 
$\nu_\mathrm{em}$ and the direction of a photon in the fluid frame is known,
its frequency in the Boyer-Lindquist coordinate frame $\nu_\mathrm{obs}$
can be calculated and this is the frequency that would be measured by 
a distant observer, unless the photon goes under the horizon. For photons
going to infinity the redshift factor can be defined:
\begin{equation}
1+z={\nu_\mathrm{em} \over \nu_\mathrm{obs}}
\end{equation}
Distant observers measure photon energies divided by the factor $(1+z)$.
Also the time interval between the detection of two signals is the interval
between their sending multiplied by $(1+z)$.
Thus the contribution to the total luminosity from the fluid element of the
volume $\Delta V$,  measured by distant
observers in their frequency interval $\Delta\nu_\mathrm{obs}$ is given as:
\begin{equation}
L_(\nu_\mathrm{obs})\Delta\nu_\mathrm{obs}=
{1 \over N}~\sum_{i=1}^N~~
{\epsilon(\nu_\mathrm{em})\Delta\nu_\mathrm{em}
~\mathrm{e}^{-\tau_{\nu_\mathrm{em}}^{(i)}}
\over (1+z_i)^2}~~\Delta V
\end{equation}
where for the i-th ray:
\begin{equation}
\nu_\mathrm{em}=\nu_\mathrm{obs}(1+z_i)~~~\Delta\nu_\mathrm{em}=
\Delta\nu_\mathrm{obs}(1+z_i)
\end{equation}
The summation is limited to rays which reach infinity. Integration over 
the volume of the fluid gives the total luminosity of the model.

The emission from the configuration is not isotropic, so the observers 
at different position angles measure a different flux of radiation. 
The luminosity calculated above is in fact an average of luminosities 
assigned to the disk by observers uniformly distributed on a sphere 
around the object. One can also find the average luminosity that would 
be measured by observers from a limited solid angle $\Delta\Omega$. 
To do so it is sufficient to neglect in the summation all the rays 
which do not enter the region of interest and multiply the result 
by the correction factor $4\pi/\Delta\Omega$.

\subsection{Spatial distribution of the photon emissivity}

In the Monte Carlo simulations we follow individual photons as they travel
through the fluid undergoing consecutive scatterings. We have to know what
is the distribution of the points of emission of the photons. We divide the
flow into several spherical layers. The radius in the middle of the
layer numbered $j$ is $r_j$. Each layer is then subdivided
into annuli of limited range in the polar angle $\theta$ between fluid
boundaries 
$\theta_\mathrm{min}(r_j)$ and $\pi-\theta_\mathrm{min}(r_j)$.
The angular coordinate at the middle of the annulus numbered $jk$ 
is $\theta_{jk}$.
One can assume that all fluid parameters are almost uniform inside each
annulus, and take their values at $(r_j,\theta_{jk})$ as representative.
Using similar arguments as in the previous subsection we calculate the
rate of photon emission from the region numbered $jk$  
of the volume $\Delta V_{jk}$:
\begin{equation}
{\dot {\cal N}}_{jk}=
{\Delta V_{jk} \over N}~\sum_{i=1}^N~~
\int_0^\infty{\epsilon(\nu_\mathrm{em})
~\mathrm{e}^{-\tau_{\nu_\mathrm{em}}^{(i)}}~\mathrm{d}\nu_\mathrm{em}
\over h\nu_\mathrm{em}(1+z_i)}
\label{ndotjk}
\end{equation}
We adopt the expressions for synchrotron cooling
from Pacholczyk (\cite{pacholczyk}) 
and Mahadevan, Narayan \& Yi (\cite{MNY}), 
and for brems\-strahlung cooling from Svensson (\cite{svensson}) [and
references therein]. 
The cooling rates are functions of the electron temperature, 
the number density of ions and electrons and the magnetic field density 
(synchrotron radiation) and hence they are functions of $r$ and $\theta$.
We take $\epsilon=\epsilon(r_j,\theta_{jk})$.
The expression under the integral is regular at low frequencies despite 
the presence of $\nu_\mathrm{em}$ in the denominator because 
$\tau_\mathrm{em}^{(i)} \rightarrow \infty$ when $\nu_\mathrm{em}
\rightarrow 0$. The redshift factor in the denominator takes care of the
difference between clock rates in the fluid frame and at infinity.

\section{The Comptonization}

We follow the method of Comp\-to\-ni\-za\-tion described by G\'{o}recki \&
Wilczewski (\cite{gorecki}).

\subsection{Basic concepts}

The differential cross section for Compton scattering is given by the
following formula (Akhiezer \& Berestetski \cite{akhiezer}):
\begin{equation}
{\mathrm{d}\sigma \over \mathrm{d}\Omega^\prime}={r^2_0 \over
2\gamma^2}X(1-\vec{v}\vec{\Omega}/c)^{-2}
{\left( {h\nu^\prime \over h\nu} \right)}^2
\end{equation}
where $h\nu$, $\vec{\Omega}$, $h\nu^\prime$ and $\vec{\Omega}^\prime$ 
are respectively
the energy and the direction of the photon before and after the scattering,
$\vec{v}$ is the velocity of the electron, $\gamma$ is the Lorentz factor and
$r_0$ is the classical electron radius. The symbol $X$ denotes  the
invariant part of the cross section :
\begin{equation}
X={x \over x^\prime}+{x \over x}^\prime+4\left({1 \over x}-{1 \over
x^\prime}\right)+4{\left({1 \over x}-{1 \over x^\prime}\right)}^2
\end{equation}
where
\begin{equation}
{x \over 2}={h\nu \over mc^2}\gamma(1-\vec{v}\vec{\Omega}/c) ~,~ 
{x \over 2}^\prime={h\nu^\prime \over 
mc^2}\gamma(1-\vec{v}\vec{\Omega}^\prime/c)
\end{equation}
are the energies of the incoming photon and of the scattered photon,
respectively, expressed in units of $mc^2$ in the reference frame of the
electron. The energies $h\nu$ and $h\nu^\prime$ are related by the
Compton formula
\begin{equation}
h\nu^\prime={h\nu(1-\vec{v}\vec{\Omega}/c) \over 
1-\vec{v}\vec{\Omega}^\prime/c+{h\nu \over 
\gamma mc^2}(1-\vec{\Omega} \vec{\Omega}^\prime)}
\label{compton}
\end{equation}
We use the total Compton cross section which is given by (Berestetski,
Lifshitz \& Pitaevski \cite{berestetski})
\begin{eqnarray}
\lefteqn{\sigma (x) =} \nonumber \\
 & = 2\pi r^2_0{1 \over x} \left[\left(1-{4 \over x}-{8 \over
x^2}\right)ln(1+x)+{1 \over 2}+{8 \over x}-{1 \over 2(1+x)^2}\right]
\end{eqnarray}
The basic concept of this method is to follow the photon trajectory from
the moment of emission until the photon leaves the flow. The probability
that a photon leaves the flow without scattering is
\begin{equation}
P_i=exp\left\{-\int_{\vec{r}_i}^{\vec{r}_\infty} 
n_e\langle\sigma\rangle \mathrm{d}l\right\}
\end{equation}
where the integral is taken along the photon trajectory from the point of
the last ($i$-th) scattering to the boundary of the flow $\vec{r}_\infty$,
$n_e=\int n_e(\vec{v})\mathrm{d}^3v$ is the electron density, 
$n_e(\vec{v})$ is the electron velocity distribution, and 
\begin{equation}
\langle \sigma \rangle = {1 \over n_e}\int
n_e(\vec{v})(1-\vec{v}\vec{\Omega}/c)
\sigma(x)\mathrm{d}^3v
\end{equation}
is a mean cross section averaged over the electron velocity distribution.
$n_e(\vec{v})$. 

The probability $P_i$ enables us to find the statistical weights of 
the number of photons leaving the flow without scattering (and thus
contributing to the emerging spectrum) and the photons which remain in
the flow and undergo the next ($[i+1]$-th) scattering. These weights are
given by $w_i P_i$ and $w_{i+1}=w_i (1-P_i)$, respectively, where
$i=0$, $1$, $2$, $3$, $...$ is the index denoting the succeeding 
scatterings. We
assume $w_0=1$. We follow the trajectory of the photon until $w$ becomes
less than a certain minimal value $w_\mathrm{min}$. Since ADAFs are optically
thin (in our model the Thomson optical depth is about 0.1 in
equatorial directions) the mean number of scatterings is 4 - 5 for 
$w_\mathrm{min}=10^{-7}$.

\subsection{Generating the random variables}

The random variables are generated from the probability distributions
using Monte Carlo methods: the inversion of the cumulative distribution
function or von Neumann's rejection technique. Multi-dimensional
distributions are modeled using the conditional probability
distributions. 

(i) At first we generate the position vector at which the
photon is initially emitted. 
According to our approximation this  position is uniformly distributed
in space within each of the annuli $jk$. The probability that a 
photon is emitted from a region of the given number $jk$ is :
\begin{equation}
f_{jk}=\frac{{\dot {\cal N}}_{jk}}
{\sum_{j^\prime,k^\prime} {\dot {\cal N}}_{{j^\prime}{k^\prime}} }
\label{efjotka}
\end{equation}
where ${\dot {\cal N}}_{jk}$ are given in Eq.~\ref{ndotjk}. In
Eq.~\ref{ndotjk} we take into account the optical depth due to the
absorption, so the number of the low frequency photons we use in the 
simulation
is the expected number of the photons which have a chance of escape. The
absorption does not have to be considered on the further photon
trajectory. (After a scattering with relativistic electrons a photon gains
so much energy that the possibility of its absorption can be
neglected. The probability of absorption on the original trajectory is
included in Eq.~\ref{ndotjk}.)

(ii) As the position of input photon is determined we can generate 
the initial energy of the photon $h\nu_0$ using a probability distribution
specified by the photon spectrum of synchrotron or brems\-strahlung emission 
\begin{equation}
f_{jk}(\nu)=\frac{n_{\nu}(r_j,\theta_{jk})}
{\int_0^{\infty}n_{\nu}(r_j,\theta_{jk})\mathrm{d}\nu}
\end{equation}   
where $n_{\nu}(r_j,\theta_{jk}) [Hz^{-1}cm^{-3} s^{-1}]$ is the photon spectrum 
approximated by formulae of Pacholczyk (\cite{pacholczyk}) 
and Mahadevan, Narayan \& Yi (\cite{MNY}) for synchrotron emission 
or Svensson (\cite{svensson}) for brems\-strahlung
emission. The photon spectrum is determined from the energy spectrum
by dividing the last one by $h\nu$.

(iii) We assume that the emission of input photons is isotropic. Hence 
the direction of the photon in a comoving Cartesian coordinate frame 
$\vec{\Omega}$= $(\sin\Theta\cos\Phi$, $\sin\Theta\sin\Phi$, $\cos\Theta)$
is generated
from uniform distributions in the ranges $\cos\Theta\in[-1;1]$ and
$\Phi\in[0;2\pi]$.

In this way we determine the set of parameters
\{$\vec{r}_0$, $h\nu_0$, $\vec{\Omega}_0$, $w_0=1$\} 
describing the initial point of
the trajectory of the photons beam. We calculate the next points of the
trajectory, i.e. 
\{$\vec{r}_{i+1}$, $h\nu_{i+1}$, $\vec{\Omega}_{i+1}$, $w_{i+1}$\} 
($i=0$, $1$, $2$, $3$, $...$) until $w=w_{min}$. The way of computing 
the weights $w_i$ is described in Sec.~3.1. Below we present the way of
computing the position, energy and direction of a photon after following
scatterings. 

(iv) The position $\vec{r}_{i+1}$ is found on the photon trajectory 
at the proper distance $l$ from the starting point
$\vec{r}_i$, from the probability distribution:
\begin{equation}
f(l)=
\frac{\mathrm{e}^{-\tau(l)}\frac{\mathrm{d}\tau(l)}{\mathrm{d}l}}
{\int_0^{l_\infty}e^{-\tau(l)}\frac{\mathrm{d}\tau(l)}
{\mathrm{d}l}\mathrm{d}l}
\end{equation}
where
\begin{equation}
\tau(l)=\int_0^{l} n_\mathrm{e}{\langle \sigma \rangle}\mathrm{d}l
\label{tauscat}
\end{equation}

(v) The two remaining parameters $h\nu_{i+1}$ and $\vec{\Omega}_{i+1}$ 
of the $(i+1)-th$ point of the photon trajectory are obtained by simulating
the scattering of the photon of energy $h\nu_i$ and direction
$\vec{\Omega}_i$ by an electron with velocity $\vec{v}$. To describe the
probability distribution of this scattering we use the differential
cross section (1) :
\begin{eqnarray}
f_{(h\nu_i,\vec{\Omega}_i)}(\vec{v},\vec{\Omega}_{i+1}) & = & 
\nonumber \\
 & = & 
\frac{n(\vec{v})(1-\vec{v}\vec{\Omega}_i/c)\frac{\mathrm{d}\sigma}{\mathrm{d}\Omega_{i+1}}}
{\int\!\!\int 
n(\vec{v})(1-\vec{v}\vec{\Omega}_i/c) 
\frac{\mathrm{d}\sigma}{\mathrm{d}\Omega^{\prime}}
\mathrm{d}^3\Omega^{\prime}\mathrm{d}^3\vec{v}}
\end{eqnarray}
We model the multi-dimensional probability distribution (13) as a
product of the probability distribution of $\vec{v}$ and the
conditional probability distribution of $\vec{\Omega}_{i+1}$
\begin{equation}
f_{(h\nu_i,\vec{\Omega}_i)}(\vec{v},\vec{\Omega}_{i+1})=
f_1(\vec{v})f_2(\vec{\Omega}_{i+1}|\vec{v})
\end{equation}
where
\begin{equation}
f_1(\vec{v})=\frac{n(\vec{v})}{n_e}(1-\vec{v}\vec{\Omega}_i/c)
\frac{\sigma(x)}{\langle \sigma \rangle}
\label{velocity}
\end{equation}
and
\begin{equation}
f_2(\vec{\Omega}_{i+1}|\vec{v})=\frac{1}{\sigma(x)}
\frac{\mathrm{d}\sigma}{\mathrm{d}\Omega_{i+1}}
\label{direction}
\end{equation}
We generate the velocity $\vec{v}$ from Eq.~\ref{velocity} 
and then the direction $\vec{\Omega}_{i+1}$ from Eq.~\ref{direction}. 
Having $\vec{\Omega}_{i+1}$ , the energy
of the scattered photon $h\nu_{i+1}$ can be obtained from the Compton
formula (\ref{compton}). The detailed description of the method of 
modeling the 
probability distributions $f_1(\vec{v})$ and 
$f_2(\vec{\Omega}_{i+1}|\vec{v})$ can be found in G\'orecki \&
Wilczewski (\cite{gorecki}).

\section{Results}

\begin{figure}
  \resizebox{\hsize}{!}{\includegraphics{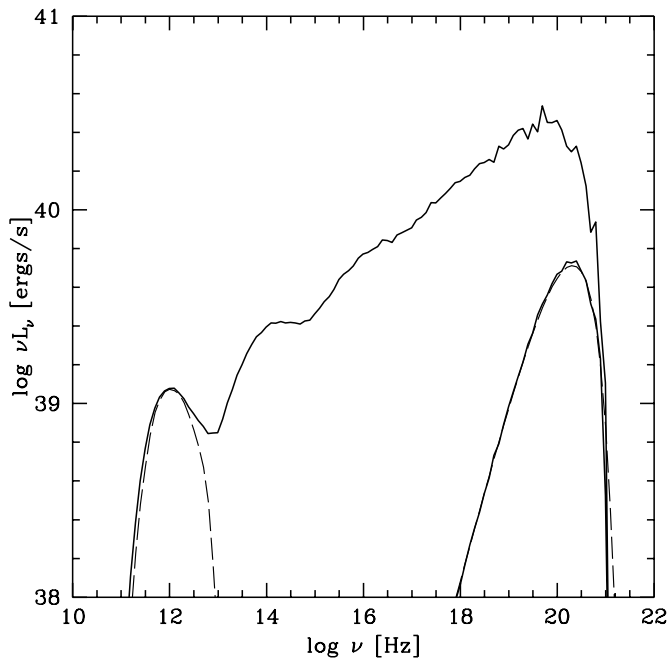}}
  \caption{
	The spectra of the input synchrotron and brems\-strahlung photons
	and the resulting spectrum after Comp\-to\-ni\-za\-tion for $a=0.9$. 
	The resulting spectrum is shown as a solid line. Synchrotron 
	input (left) and brems\-strahlung (right) use dotted lines.
	}
  \label{spectrum}
\end{figure}

We have performed trial calculations of the ADAFs spectra employing the
method described. We use the models of ADAFs from Paper I. For the black hole
mass, accretion rate and parameter $\beta$ we use the parameters of
Lasota et al. (\cite{jpl96}), 
$M_\mathrm{BH}=3.6 \times 10^7 M_{\sun}$, ${\dot m}=0.016$, $\beta=0.95$,
which they apply
in the modeling of \object{NGC 4258}. We do not, however, attach  
cold, thin
disks at large radii to our ADAF solutions.
In Fig.~\ref{spectrum} we show the input spectra of
synchrotron and brems\-strahlung photons as well as the resulting
Comptonized spectrum of the disk around $a=0.9$ black hole. 
All the spectra in this and other diagrams  are shown as $\lg(\nu)$ versus
$\lg(\nu F_\nu)$ plots.
For the synchrotron input the results
are based on calculations including $10^6$ input photons and
following more than $5\times 10^6$ branches of photon trajectories. The
Comp\-to\-ni\-za\-tion plays a less important role in the case of brems\-strahlung
radiation, so we  use $\sim 10$ times fewer photon trajectories to
obtain the spectra in this case. Since the Comp\-to\-ni\-za\-tion preserves the
photon number, we normalize the spectra using Eq.~\ref{ndotjk} with
either synchrotron or brems\-strahlung emissivity under the integral to
obtain the relative numbers of seed photons of each kind.

\begin{figure}
  \resizebox{\hsize}{!}{\includegraphics{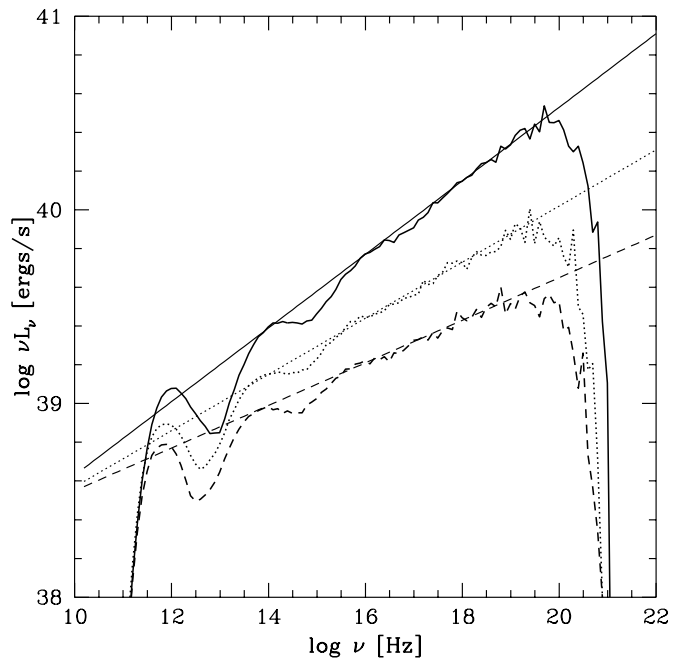}}
  \caption{
	The power law fits to the Comptonized synchrotron emission for
	models with black hole angular momentum
	$a=0$ (dashed lines), $a=0.5$ (dotted) and $a=0.9$ (solid).
	The corresponding straight lines represent the fits.
	}
  \label{fits}
\end{figure}

The Comptonized spectra of synchrotron radiation are smooth enough to
allow a power law fit with power index $\Gamma$ (i.e. $L_\nu \sim
\nu^{-\Gamma}$). The fit is not valid at the vicinity of the first peak, 
which is due to the seed photons. In Fig.~\ref{fits} we show synchrotron
spectra with fits. As can be seen in the plots the slopes of the
spectra depend on the model. Since the black hole mass and the accretion
rate are the same for all three cases, the differences must be attributed
to the black hole angular momentum and its influence on the flow structure.  
The power law indices estimated from fits
are $\Gamma=0.89$, $0.85$, and $0.81$ for $a=0.$, $0.5$, and
$0.9$ respectively. 

\begin{figure}
  \resizebox{\hsize}{!}{\includegraphics{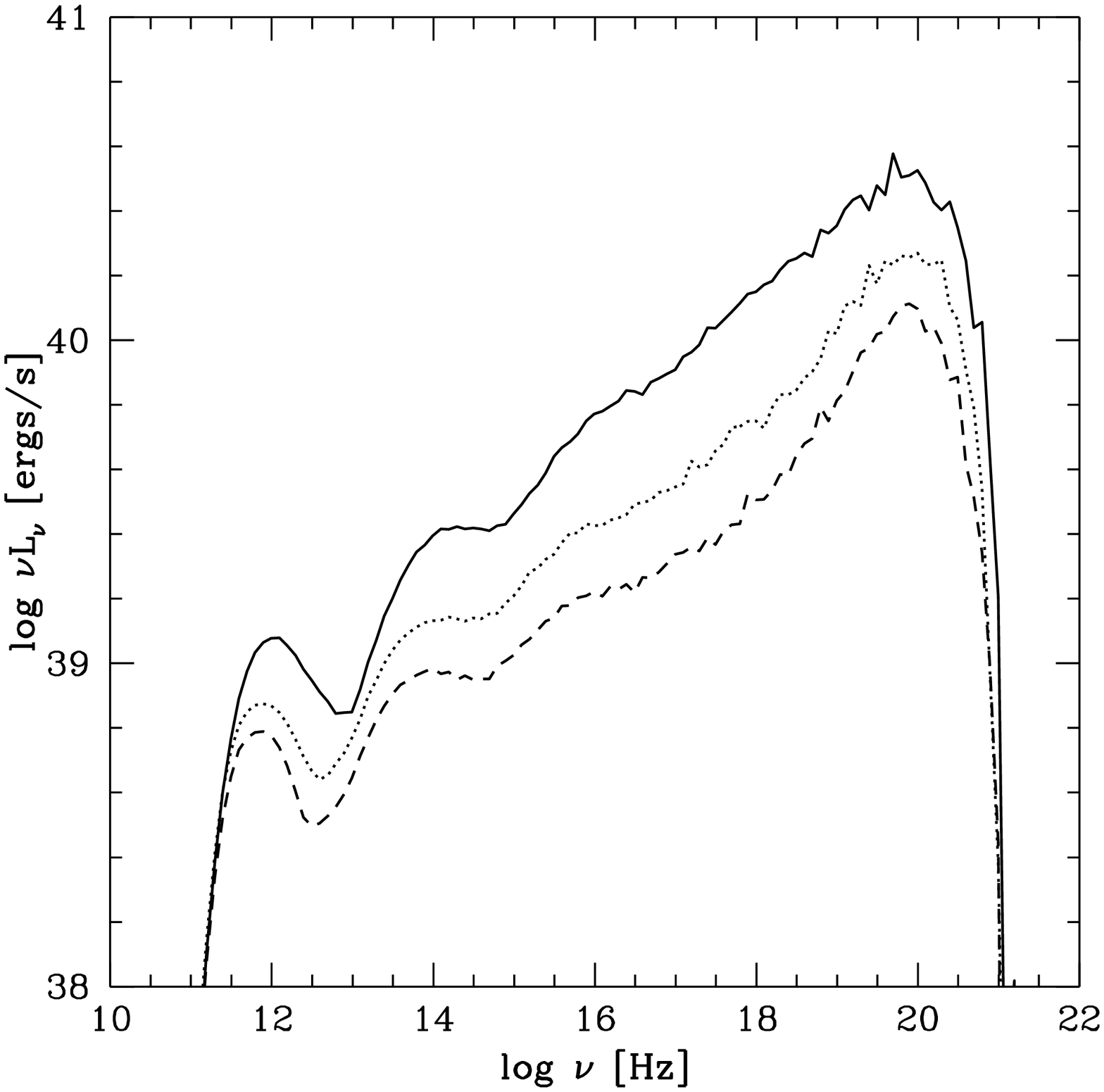}}
  \caption{
	The resulting spectra of ADAFs for the three cases including the
	brems\-strahlung component. The conventions follow Fig.~\ref{fits}
		}
  \label{widma}
\end{figure}

The three spectra resulting from combined effects of synchrotron and
brems\-strahlung emission with Comp\-to\-ni\-za\-tion are shown in 
Fig.~\ref{widma}.  

We have also checked the dependence of the observed total luminosity of our
models on the observer's position. For the brems\-strahlung photons the
dependence is absent. The synchrotron radiation observed from the
equatorial plane is stronger by 10 to 20\% as compared to the measurement
from the axis of rotation.

Following the photons we are able to find the fraction which goes under the
horizon. For the synchrotron photons the numbers are: $0.07$, $0.05$, and
$0.04$ for $a=0$, $0.5$ and $0.9$ respectively. The brems\-strahlung photons
are emitted at relatively larger distances from the horizon and less than
1\% of them are lost in all cases.
The fraction of photons emitted by the fluid and going under the horizon
is a decreasing function of the black hole angular momentum according to
our simulations. 
We have checked this result of our simulations making an independent
calculation. We have compared isotropic sources of radiation at the
same distance from the black hole, comoving with the matter of the three
models we use. Again the fraction of rays going under the horizon is the
smallest for the case of the most rapidly rotating hole. The effect must
be attributed to the differences in matter kinematics between the three
models. 

\section{Discussion and conclusions}

We use the ADAF models from Paper I, but we apply some changes to the
treatment of the 3D structure of the flow. The models are still based on
vertically averaged, stationary equations for the radial structure of the
flow, which uses the vertical scale height and quantities measured at
the equator as flow variables. Such treatment can be inadequate near 
the flow boundary, especially close to the rotation axis. We assume the
specific angular momentum of matter, its radial velocity, and the sound
speed to be constant on spheres. Under such assumptions it is
possible to introduce an effective potential on each sphere, which
acts as infinite centrifugal potential barrier near the rotation axis.
This infinite barrier causes the sharp drop to zero of matter density
there despite the fact that the gas is approximately isothermal on spheres;
it effectively removes matter with artificially high angular velocity.

Our models are based on the assumption that all the heat dissipated in the
flow goes into the ions. We neglect the direct viscous heating of electrons
and the fact that their entropy changes as they fall toward the black hole, 
which represents the advection of heat.  Both mechanisms influence the
energy balance equation and,  as shown by Nakamura et
al. (\cite{nakamura97}), Narayan et  al. (\cite{narayan98}) and Quataert \&
Narayan (\cite{QN}), have an impact on the electron temperature and the
resulting spectrum of the model. While advection of heat by electrons is a
well defined process, the viscous heating must be introduced using another
free parameter. The combined effects of viscosity and advection on
electrons would influence all our models in a similar way, not greatly
changing the differences between them.

In our calculations we use three models of the matter flow onto the black
hole, with the same accretion rate and the same black hole mass, but with
different black hole angular momentum. The spin of the black hole has
strong influence on the density and temperature of the matter near the
horizon, which are both increasing functions of the rotation rate.
Similar behavior of the gas parameters can also be seen in much broader
investigation of ADAFs parameter space by Popham \& Gammie
(\cite{popham98}). We are not able to present a full discussion of the
ADAF structure - spectrum dependence, but we can point out some trends.

Our calculations show a strong dependence of the ADAFs spectra on the
flow structure resulting from the differences in the black hole angular
momentum. The synchrotron seed photons are produced mainly in the
central parts of the flow, which are the densest and the hottest. The
total energy emitted as  synchrotron photons increases with the black
hole angular momentum. Also the influence of Comp\-to\-ni\-za\-tion is
increased the same way. In the case of $a=0.9$ model, the Comptonized
synchrotron radiation dominates all the way to the highest frequencies,
making the usual brems\-strahlung peak invisible. For other cases
considered ($a=0.5$ or $0$) this is not true and the brems\-strahlung
components dominate at highest frequencies. 

The standard theory of Comp\-to\-ni\-za\-tion (Rybicki \& Lightman 
\cite{rybicki})
enables one to estimate the spectral index of low energy radiation,
which undergoes multiple scatterings with thermal relativistic
electrons of given temperature and optical depth. In our case both
parameters can be defined as averages over the configuration. We have
tried several simple prescriptions for calculating the averages, but we
have not obtained a quantitative agreement between our results and the
estimates, the calculated spectra having spectral indexes by $\approx
0.1$ higher (i.e. being steeper). The discrepancies must be attributed
to the complexity of the flow and effects such as the relative motion of
the starting point of a photon and the place of its interaction with the
electrons. 

We are not attempting to model \object{NGC 4258}, but using the
parameters from the model of Lasota et al. (\cite{jpl96}) we get the
right luminosity in the X-rays for the $a=0.9$ model. The slope of the
calculated spectrum at this frequency ($10^{18}~\mathrm{Hz}$) is  within the
observational bounds. The spectrum of \object{NGC 4258} has been also 
modeled  by Gammie, Narayan, \& Blandford (\cite{gnb}). 
Although they use slightly different values of the ADAF parameters, 
their results are very similar to those of Lasota et al. (\cite{jpl96}).

\begin{acknowledgements}
  This work was supported in part by the Polish State Committee
for Scientific Research grant 2-P03D-012-12
\end{acknowledgements}

\end{document}